\documentclass[11pt,twoside]{article}
\usepackage{amsmath}
\usepackage{amssymb}
\usepackage{amscd}
\usepackage{mathrsfs}
\newcommand{\remark}{\smallbreak\noindent{\bf Remark.}}
\newcommand{\myitem}{\smallbreak\noindent$\bullet$}
\newcommand{\sref}[1]{\S\ref{#1}}
\newcommand{\onto}{\rightarrowtail}
\newcommand{\CC}{{\mathbb{C}}}
\newcommand{\LL}{{\mathbb{L}}}
\newcommand{\RR}{{\mathbb{R}}}
\newcommand{\Id}[1]{{1\!\!1}\!{}_{#1}{}}
\newcommand{\id}{{1\!\!1}}
\newcommand{\we}{{\,\wedge\,}}
\newcommand{\weu}[1]{{\wedge^{\!#1}}}
\newcommand{\tn}{{\,\otimes\,}}
\newcommand{\comp}{\mathbin{\raisebox{1pt}{$\scriptstyle\circ$}}}
\newcommand{\pint}{{\scriptscriptstyle\mathord{\rfloor}}}
\newcommand{\bang}[1]{{\langle#1\rangle}}
\newcommand{\fnb}[1]{[\![#1]\!]}
\newcommand{\de}{\partial}
\newcommand{\na}{\nabla\!}
\newcommand{\nasl}{{\rlap{\raise1pt\hbox{\,/}}\nabla}}
\newcommand{\Cs}{{\rlap{\lower3pt\hbox{\textnormal{\LARGE \char'040}}}{\Gamma}}{}}
\newcommand{\spec}[1]{{}_{\scriptscriptstyle{\mathrm{#1}}}}
\newcommand{\cev}[1]{\smash{\overset{\smash{{}_{\gets}}}{#1}}}
\newcommand{\ost}[1]{\overset{{}_{{\,}_*}}{#1}}
\newcommand{\G}{\Gamma}
\renewcommand{\d}{\delta}
\newcommand{\e}{\varepsilon}
\newcommand{\Th}{\Theta}
\newcommand{\bb}{{\mathsf{b}}}
\newcommand{\xx}{{\mathsf{x}}}
\newcommand{\dx}{\dO\xx}
\newcommand{\Ii}[2]{{}^{#1}_{\phantom{#1}\!#2}}
\newcommand{\iI}[2]{{}_{#1}^{\phantom{#1}\!#2}}
\newcommand{\iIi}[3]{{}_{#1\phantom{#2}\!\!#3}^{\phantom{#1}\!#2}}
\renewcommand{\.}{{\scriptstyle\boldsymbol{\dot{}}}}
\newcommand{\sA}{{\scriptscriptstyle A}}
\newcommand{\sB}{{\scriptscriptstyle B}}
\newcommand{\sC}{{\scriptscriptstyle C}}

\newcommand{\cA}{{\sA\.}}
\newcommand{\cB}{{\sB\.}}
\newcommand{\cC}{{\sC\.}}

\newcommand{\AAd}{{\sA\cA}}
\newcommand{\BBd}{{\sB\cB}}
\newcommand{\CCd}{{\sC\cC}}

\newcommand{\sX}{{\scriptstyle X}}
\newcommand{\ssX}{{\scriptscriptstyle X}}
\newcommand{\sZ}{{\scriptstyle Z}}
\newcommand{\sDe}{{\scriptstyle\Delta}}
\newcommand{\Sc}{\overline{S}}
\newcommand{\Sa}{\overline{S}{}^*}
\newcommand{\Uc}{\overline{U}}
\newcommand{\Ua}{\overline{U}{}^*}
\newcommand{\Vc}{\overline{V}}
\newcommand{\Va}{\overline{V}{}^*}
\newcommand{\Wc}{\overline{W}}

\newcommand{\dO}{\mathrm{d}}
\newcommand{\iO}{\mathrm{i}}
\newcommand{\po}{\mathrm{p}}
\newcommand{\HO}{\mathrm{H}}
\newcommand{\LO}{\mathrm{L}}
\newcommand{\TO}{\mathrm{T}}
\newcommand{\TS}{\TO^{*}\!}
\newcommand{\VO}{\mathrm{V}}
\newcommand{\End}{\operatorname{End}}
\newcommand{\Tr}{\operatorname{Tr}}
\newcommand{\oh}{\tfrac{1}{2}}
\newcommand{\oq}{\tfrac{1}{4}}
\newcommand{\oo}{\tfrac{1}{8}}
\newcommand{\osq}{\tfrac{1}{\surd2}}
\newcommand{\isq}{\tfrac{\iO}{\surd2}}
\title{Two-spinor tetrad and Lie derivatives\\
of Einstein-Cartan-Dirac fields}
\date{{\small September 28, 2016} }
\author{Daniel Canarutto \\[6pt]
{\small\it Dipartimento di Matematica e Informatica ``U.~Dini'', }\\
{\small\it Via S. Marta 3, 50139 Firenze, Italia}\\
{\small email:~daniel.canarutto@unifi.it}\\
{\small http://www.dma.unifi.it/\char126 canarutto}}
\begin{document}
\maketitle \thispagestyle{empty}

\begin{abstract}\noindent
An integrated approach to Lie derivatives of spinors,
spinor connections and the gravitational field is presented,
in the context of a previously proposed, partly original formulation
of a theory of Einstein-Cartan-Maxwell-Dirac fields
based on ``minimal geometric data'':
all the needed underlying structure is geometrically constructed
from the unique assumption of a complex vector field \hbox{$S\onto M$}
with 2-dimensional fibers.
The Lie derivatives of objects of all considered types,
with respect to a vector field \hbox{$\sX:M\to\TO M$},
are well-defined without making any special assumption about $\sX$,
and fulfill natural mutual relations.
\end{abstract}
\bigbreak
\noindent
MSC 2010:
53B05, 
58A32, 
83C60. 

\bigbreak
\noindent
PACS 2010:
02.40.-k, 
04.20.Cv. 

\bigbreak
\noindent
Keywords: Lie derivatives of spinors, Lie derivatives of spinor connections,
deformed tetrad gravity.

\bigbreak\bigbreak
\noindent

\vfill\newpage
\tableofcontents
\thispagestyle{empty}
\vfill\newpage
\setcounter{page}{1}

\section*{Introduction}

Lie derivatives of 4-spinors on curved spacetime have been studied
by Kosmann~\cite{Kosmann71}
and others~\cite{FatibeneFerrarisFrancavigliaGodina96,GodinaMatteucci05,
LeaoRodriguesWainer15}
by exploiting structure groups and their representations
in order to extend to spinors the notion of transport of tensor fields
by the local 1-parameter group associated with a vector field.
A somewhat different approach by Penrose and Rindler~\cite{PR88}
recovers the Lie derivative of a 2-spinor from the requirement
that it is related to the usual Lie derivative through the Leibnitz rule.
A recent article~\cite{Helfer16} examined the relations among various approaches.

One key point about this topic is the ``soldering'' of the spinor
and spacetime geometries.
Usually, the soldering is implicitely contained in the formalism;
we propose to make it explicit
by means of a partly original formulation of tetrad gravity,
described in previous papers~\cite{C98,C00b},
which yields an integrated treatment of Einstein-Cartan-Maxwell-Dirac fields
starting from minimal geometric data.

The tetrad formalism~\cite{Cartan1922,Cartan1923_25,Hehl1973_I,Hehl1974_II,HehlHeydeKerlickNester76,
HCMN95,Henneaux78,Poplawski09,Sciama54,Trautman06}
could be just regarded as using an orthonormal spacetime frame
in order to describe gravitation.
A geometric refinement can be achieved by assuming
a vector bundle \hbox{$H\onto M$} whose 4-dimensional fibers
are endowed with a Lorentz metric,
and defining the tetrad as a ``soldering form'' between $H$
and the tangent bundle of $M$.
This extra assumption, apparently contrary to Ockam's ``razor principle'',
can be actually turned into a free benefit since $H$ can be derived,
by a geometric construction,
from a complex vector bundle \hbox{$S\onto M$} with 2-dimensional fibers.
The same $S$ naturally yields the bundle $W$ of 4-spinors
together with the Dirac map \hbox{$H\to\End W$},
and any other structure needed for the aforementioned integrated field theory
(length units included).
The tetrad itself replaces the spacetime metric $g$,
and indeed it can be regarded as a ``square root'' of $g$,
while a connection of $S$ naturally splits into ``gravitational''
and ``electromagnetic'' contributions.
The spacetime connection, on the other hand,
is no more regarded as a fundamental field but rather as a ``byproduct''.
The underlying 2-spinor formalism is compatible
with the Penrose-Rindler formalism~\cite{PR84,PR88},
with a few adjustments.

After an essential account of the above said setting
we address the notions of Lie derivatives of all involved fields
with respect to a vector field \hbox{$\sX:M\to\TO M$}.
We find that by explicitely taking the tetrad into account
we are able to give a natural definition of Lie derivative of a spinor
of any type without imposing any constraint on $\sX$
(such as being Killing or conformal Killing).
The Lie derivative of the tetrad itself takes care of what is missing.
Furthermore we obtain a natural definition of the Lie derivative
of the spinor connection.

The notion of the Lie derivative of a linear connection
of the tangent space of a manifold,
and the related notion of deformed connection,
have been known for a long time (see e.g.\ Yano~\cite{Yano55}, \S I.4).
The main use of that notion in the literature deals with energy tensors
in General Relativity~\cite{KijowskiGRG97,Padamnabhan1312.3253},
possibly in the disguised form of ``deformations''
of the spacetime connection~\cite{LandauLifchitz68,HE73}.
Exploiting that concept, the Lie derivative of the spinor connection
can be introduced similarly to the Lie derivative of spinors,
by a procedure that uses the tetrad and the corresponding natural decompositions
of the spaces of endomorphisms of $S$ and $H$.
We examine the relations among the various considered operations
and write down the basic coordinate formulas.

Finally we discuss the notion of a ``deformed'' theory
of Einstein-Cartan-Dirac fields,
and offer remarks about deformations
of fields whose ``internal degrees of freedom''
are not soldered to spacetime geometry.

\section[Einstein-Cartan-Maxwell-Dirac fields]%
{Einstein-Cartan-Maxwell-Dirac fields using ``minimal geometric data''}
\label{s:ECMD fields}

The first three sections that follow deal with purely algebraic constructions,
whose only ingredient is a 2-dimensional complex vector space $S$.
Afterwards we'll consider a vector bundle \hbox{$S\onto M$}
over a real 4-dimensional manifold $M$,
where our constructions can be performed fiberwise
yielding various associated bundles and natural maps.
Any topological constraint needed for everything to be sufficiently regular
will be implicitely assumed to hold, without further discussion.

Most of the material of the next six sections constitutes a summary
of previous work~\cite{C98,C00b,C07}, with some adaptations.
It won't look unfamiliar to the reader who is acquainted with
the usual two-spinor formalism,
though there are a few actual differences and, moreover,
our privileging intrisic geometric arguments over
coordinate computations yields a somewhat shifted point of view.

\subsection{Two-spinor algebra and Lorentzian geometry}
\label{ss:Two-spinor algebra and Lorentzian geometry}

A complex vector space $V$ of arbitrary finite dimension yields the associated
\emph{dual space} $V^*$,
\emph{conjugate space} $\Vc$ and \emph{anti-dual space} $\Va$.
The latter can be regarded as the space
of all anti-linear maps \hbox{$V\to\CC$}
(fulfilling \hbox{$f(cv)=\bar c\,f(v)$}, \hbox{$c\in\CC$}),
and we have the natural identification \hbox{$\Vc\cong(\Va)^*$}.
Complex conjugation determines anti-isomorphisms \hbox{$V\leftrightarrow\Vc$}
and \hbox{$V^*\leftrightarrow\Va$}.
Using this together with transposition
we also obtain an antilinear involution of $V\tn\Vc$,
determining a decomposition into the direct sum
of the \emph{real} eigenspaces
corresponding to eigenvalues $\pm1$\,, namely
$$V\tn\Vc=\HO(V\tn\Vc)\oplus \iO\,\HO(V\tn\Vc)~,$$
called the \emph{Hermitian} and \emph{anti-Hermitian} subspaces, respectively.

A basis $\bigl(\bb_\alpha\bigr)$ of $V$ yields the conjugate basis 
$\bigl(\bar\bb_{\alpha\.}\bigr)$ of $\Vc$.
If \hbox{$v\in V$} has components $v^\alpha$
then its conjugate \hbox{$\bar v\in\Vc$} has components $\bar v^{\alpha\.}$.
Accordingly, Hermitian tensors \hbox{$w\in V\tn\Vc$} are characterized
by their components fulfilling \hbox{$\bar w^{\alpha\.\,\alpha}=w^{\alpha\alpha\.}$}.

The above basic construction is the source of much of the rich algebraic
structure which can be extracted
from a 2-dimensional complex vector space $S$ without any further assumption.
The relation to a more familiar formalism for most readers
is seen by using a basis $\bigl(\xi_\sA\bigr)$ of $S$.
We distinguish a few steps.

\myitem~We start by observing that the antisymmetric subspace
\hbox{$\weu2S\subset S\tn S$} is a 1-dimensional complex vector space.
The Hermitian subspace of \hbox{$\weu2S\tn\weu2\Sc$}
is a real $1$-dimensional vector space with a distinguished orientation;
its positively oriented semispace $\LL^2$
(whose elements are of the type $w\tn\bar w$\,, $w\in\weu2S$)
has the ``square root'' semispace $\LL$,
which will can be identified with the space
of \emph{length units}.\footnote{\label{footnote:unitspaces}
Essentially, $\LL$ being the square root of $\LL^2$ means
that there is a natural isomorphism \hbox{$\LL^2\cong\LL\tn\LL$}\,.
A mahematically precise treatment of unit spaces
and physical dimensions can be found in works by
Modugno and others~\cite{JMV10,C12a}.} 
In the ensuing field theory,
$\LL$ will be the natural target for the \emph{dilaton}.
The chosen basis of $S$ determines a basis in each of the associated spaces and,
in particular, a length unit \hbox{$l\in\LL$}\,.

\myitem~Rational roots of unit spaces are well-defined,
with negative exponents standing for duality.
Accordingly we introduce the new 2-dimensional space
\hbox{$U\equiv\LL^{-1/2}\tn S$},
which has the induced basis
\hbox{$\bigl(\zeta_\sA\bigr)=\bigl(l^{-1/2}\,\xi_\sA\bigr)$}.
This is our \emph{$2$-spinor space}.
Now since \hbox{$U^*=\LL^{1/2}\tn S^*$},
the 1-dimensional complex space $\weu2U$
turns out to be naturally endowed with a Hermitian metric,
namely the identity element in
$$\LL^2\tn\LL^{-2}\cong
\LL^2\tn\HO[(\weu2\Sa)\tn(\weu2S^*)]
\cong\HO[(\weu2\Ua)\tn(\weu2U^*)]~.$$
Hence any two normalised elements in $\weu2U^*$
are related by a phase factor.
The chosen basis determines one such element,
namely \hbox{$\e=\e_{\sA\sB}\,\zeta^\sA\we\zeta^\sB$},
where \hbox{$\e_{\sA\sB}=\d^1_\sA\,\d^2_\sB-\d^1_\sB\,\d^2_\sA$}
are the antisymmetric Ricci constants and $\bigl(\zeta^\sA\bigr)$
denotes the dual basis of $U^*$.
Each normalised \hbox{$\e\in\weu2U^*$} yields the isomorphism
\hbox{$\e^\flat:U\to U^*:u\mapsto u^\flat:=\e(u,\_)$}\,,
with the coordinate expression
\hbox{$u_\sB\equiv(u^\flat)_\sB=\e_{\sA\sB}\,u^\sA$}\,.
The dual construction also yields \hbox{$\e^\#=\e^{\sA\sB}\,\zeta_\sA\we\zeta_\sB$}\,,
which yields the inverse isomorphism \hbox{$U^*\to U$}.

\myitem~We'll be specially involved with the Hermitian subspace
$$H\equiv\HO(U\tn\Uc)\subset U\tn\Uc~,$$
a 4-dimensional real vector space which turns out to be naturally endowed
with a Lorentz metric.
Actually if \hbox{$\e\in\weu2U^*$}
is normalized then \hbox{$\e\tn\bar\e\in\weu2U^*\tn\weu2\Ua$}
is independent of the phase factor in $\e$\,,
thus it is a natural object,
which can be seen as a bilinear form $g$ on $U\tn\Uc$ via the rule
$$g(u\tn\bar v,r\tn\bar s)=\e(u,r)\,\bar\e(\bar v,\bar s)=
\e_{\sA\sB}\,\bar\e_{\cA\cB}\,u^\sA r^\sB \bar v^\cA \bar s^\cB~.$$
Now if $\bigl(\bar\zeta_\cA\bigr)$ denotes the induced ``conjugate'' basis of $\Uc$
then the induced basis of $U\tn\Uc$ is $\bigl(\zeta_\sA\tn\bar\zeta_\cA\bigr)$;
alongside with this we also consider the basis $\bigl(\tau_\lambda\bigr)$,
defined in term of the Pauli matrices as
$$\tau_\lambda\equiv\osq\,\sigma_\lambda^\AAd\,\zeta_\sA\tn\bar\zeta_\cA~,
\qquad \lambda=0,1,2,3.$$
A straightforward computation then shows that this is an orthonormal
basis of $H$, with squares $(+1,-1,-1,-1)$.
Null elements in $H$ are of the form $\pm u\tn\bar u$ with \hbox{$u\in U$}
(thus there is a distinguished time-orientation in $H$).

\subsection{Two-spinors and Dirac spinors}
\label{ss:Two-spinors and Dirac spinors}

Our step-by-step constructions continue by considering the 4-dimensional
complex vector space \hbox{$W\equiv U\oplus\Ua$}.
This can be naturally regarded as the space of 4-spinors,
as we can exhibit a natural linear map \hbox{$\gamma:U\tn\Uc\to\End(W)$}
whose restriction to the Minkowski space $H$ turns out to be a Clifford map.
It is characterized by
$$\gamma(r\tn\bar s)(u,\chi)=
\sqrt2\bigl(\bang{\bar\lambda,\bar s}\,p\,,
\bang{r^\flat,u}\,\bar s^\flat\,\bigr)~,\qquad
u,p,r,s\in U\,,~\chi\in\Ua\,,$$
an expression which is independent of the phase factor in the normalized $\e$
yielding \hbox{$r^\flat\in U^*$} and \hbox{$\bar s^\flat\in\Ua$}.
The usual Weyl representation can be recovered by using the basis
$$\bigl(\omega_\alpha\bigr)\equiv
(\zeta_1\,,\zeta_2\,,-\bar\zeta^1,-\bar\zeta^2)~,$$
where $\zeta_1$ is a simplified notation for $(\zeta_1\,,0)$, and the like:
setting \hbox{$\gamma_\lambda\equiv\gamma(\tau_\lambda)\in\End W$}\,,
\hbox{$\lambda=0,1,2,3$},
the matrices $\bigl(\gamma\iIi\lambda\alpha\beta\bigr)$ in this basis
turn out to be the Weyl matrices.
By a suitable basis transformation one also recovers the Dirac representation.

Next we observe that the conjugate space of $W$ is \hbox{$\Wc=\Uc\oplus U^*$},
whence by inverting the order of the two sectors we obtain the dual space
\hbox{$U^*\oplus\Uc=W^*$}.
Let's explicitely denote this switching map,
which is obviously an isomorphism, as
$$\mathrm{s}:\Wc\to W^*:(\bar u,\lambda)\mapsto(\lambda,\bar u)~.$$
If \hbox{$\psi\equiv(u,\bar\lambda)\in W$}
then applying the conjugation anti-isomorphism to it we get
$$\bar\psi=(\bar u,\lambda)\in\Wc \quad\Rightarrow\quad
\mathrm{s}(\bar\psi)\in W^*~.$$
This $\mathrm{s}(\bar\psi)$ is exactly the object
which is traditionally denoted as $\bar\psi$\,.
When no confusion arises, we may as well adopt that notation as a shorthand.
Note that the mapping \hbox{$\psi\mapsto\bar\psi$},
called \emph{Dirac adjunction},
can be regarded as associated with a Hermitian scalar product on $W$,
which turns out to have signature $(++--)$,
as one sees immediately in the Dirac representation.

We end this section with a few remarks about aspects of our presentation
which are different from the usual 2-spinor and 4-spinor formalisms.

\myitem~No complex symplectic form is fixed.
The  $2$-form $\e$ is unique up to a phase factor
which depends on the chosen 2-spinor basis,
and yields the isomorphisms $\e^\flat$ and $\e^\#$\,.

\myitem~No Hermitian scalar product on $S$ or $U$ is assigned;
the choice of a positive Hermitian scalar product on $U$
essentially amounts to the choice of an ``observer''
in the Minkowski space $H$.

\myitem~Consequently there is no fixed complex symplectic form
nor positive Hermitian structure on the 4-spinor space $W$ as well.
The usual mapping \hbox{$\psi\mapsto\psi^\dagger$} is related to
a positive Hermitian structure associated with an observer
(while Dirac adjunction is observer-independent).
Charge conjugation is related to the choice of $\e$
(namely of a phase factor).
\smallbreak

More generally, we may observe that matrix-based formalisms tend to screen
the precise role of the various involved objects by apparently putting
different operations on the same footing.

\subsection{Endomorphism decomposition in spinor and Minkowski spaces}
\label{ss:Endomorphism decomposition in spinor and Minkowski spaces}

In the vector space \hbox{$\End H\equiv H\tn H^*$} of all
linear endomorphisms of $H$
we have Lorentz metric transposition
\hbox{$\End H\to\End H:K\mapsto K^\dagger$},
where \hbox{$(K^\dagger)\Ii\lambda\mu=
g^{\lambda\nu}\,K\Ii\rho\nu\,g_{\rho\mu}$}\,.
The subspace of all endomorphisms which are antisymmetric with respect
to this operation is the Lie subalgebra\footnote{
$\End H$ together with the ordinary commutator is a Lie algebra.} 
$\mathfrak{so}(H,g)$\,.
We obtain a natural decomposition
$$\End H=\mathfrak{so}(H,g)\oplus\RR\Id{H}\oplus\mathcal{S}_0H~,$$
where $\RR\Id{H}$ is the subspace generated by the identity of $H$
and $\mathcal{S}_0H$ is the space of all trace-free symmetric endomorphisms.
Indeed, any \hbox{$K\in\End H$} can be uniquely decomposed as
$$K=\oh\,\bigl(K\,{-}\,K^\dagger\bigr)+\oq\Tr\!K\,\Id{H}
+\bigl(\oh\,(K\,{+}\,K^\dagger)-\oq\Tr\!K\,\Id{H}\bigr)~.$$
In particular we have a projection
$$\po:\End H\to\mathfrak{so}(H,g)\oplus\RR\Id{H}:
K\mapsto\oh\,\bigl(K\,{-}\,K^\dagger\bigr)+\oq\Tr\!K\,\Id{H}~,$$
whose target space is a Lie-subalgebra of $\End H$
(while its complementary space $\mathcal{S}_0H$
is not closed with respect to the commutator).

Similarly, the vector space \hbox{$\End U\equiv U\tn U^*$} of all
$\CC$-linear endomorphisms of $U$ has the natural decomposition
$$\End U=\mathfrak{sl}(U)\oplus\CC\Id{U}=
\mathfrak{sl}(U)\oplus\RR\Id{U}\oplus\iO\RR\Id{U}$$
(where $\mathfrak{sl}(U)$ is the Lie subalgebra of all trace-free endomorphisms),
as any \hbox{$k\in\End U$} can be uniquely decomposed as
\hbox{$k=(k-\oh\Tr k\,\Id{U})+\oh\Tr k\,\Id{U}$}\,,
and the trace can be further decomposed into its real and imaginary parts.
Now recalling \hbox{$H\subset U\tn\Uc$} we introduce $\RR$-linear maps
\hbox{$\pi:\End H\to\End U$} and \hbox{$\imath:\End U\to\End H$} as follows.
The former is defined via traces and can be best expressed in component form as
$$(\pi K)\Ii\sA\sB=
\oh\,K\Ii{\sA\cA}{\sB\cA}-\oo\,K\Ii\CCd\CCd\,\d\Ii\sA\sB~.$$
The latter is defined as \hbox{$\imath k\equiv k\tn\bar\id+\id\tn\bar k$},
and is expressed in component form as
$$(\imath k)\Ii\AAd\BBd=
k\Ii\sA\sB\,\d\Ii\cA\cB+\d\Ii\sA\sB\,\bar k\Ii\cA\cB~.$$

The following statements are then easily checked:

\myitem~\hbox{$\pi\Id{H}=\oh\,\Id{U}$}\,, \hbox{$\imath\Id{U}=2\,\Id{H}$}\,.

\myitem~$\pi$ and $\po$ have the same kernel: the symmetric traceless sector
of $\End H$.

\myitem~The kernel of $\imath$ is the imaginary part of the identity sector.

\myitem~Both $\po$ and $\imath$ are valued onto
\hbox{$\mathfrak{so}(H,g)\oplus\RR\Id{H}$}\,.

\myitem~The restriction of $\pi$
to \hbox{$\mathfrak{so}(H,g)\oplus\RR\Id{H}$}
and the restriction of $\imath$ to \hbox{$\mathfrak{sl}(U)\oplus\RR\Id{U}$}
are inverse Lie-algebra isomorphisms.
\smallbreak

\remark~The definition of $\imath$ is crafted in such a way
that the action of $\imath k$ on isotropic elements \hbox{$u\tn\bar u\in H$}
is determined by the Leibnitz rule
\hbox{$\imath k(u\tn\bar u)=ku\tn\bar u+u\tn\bar k\bar u$}\,.
This is perhaps the most relevant aspect of this matter in relation to
Lie derivatives of spinors.
Also note that this expression is closely related to the decomposition
\hbox{$\Phi_{\AAd\BBd}=
\phi_{\sA\sB}\,\bar\e_{\cA\cB}+\e_{\sA\sB}\,\bar\phi_{\cA\cB}$}\,,
valid for a Minkowski space 2-tensor $\Phi$ whose symmetric part
is proportional to the Lorentz metric.\smallbreak

We now introduce a further map, the $\RR$-linear inclusion\footnote{
A detailed examination in two-spinor terms
of Lie groups and Lie algebras involved in spinor and Minkowsky space
geometries can be found in a previous work~\cite{C14}.} 
$$\varkappa:\End U\to\End W:k\mapsto\varkappa(k)\equiv(k,-\bar k^*)~,$$
where \hbox{$\bar k^*:\Ua\to\Ua$} is the conjugate transpose of $k$\,.

The composition \hbox{$\pi\comp\varkappa:\End H\to\End W$}
can be then expressed in terms of the components
of the Dirac map as
$$\varkappa(\pi K)=
\oo K_{\lambda\mu}\,(\gamma^\lambda\gamma^\mu-\gamma^\mu\gamma^\lambda)
+\oo K\Ii\nu\nu\,\gamma_5~,$$
where \hbox{$\iO\,\gamma_5\equiv\gamma_0\gamma_1\gamma_2\gamma_3$}
is the element in the Dirac algebra
corresponding to the natural volume form of $H$.

A diagram of the mutual relations among the introduced maps
may be useful:
$$\begin{picture}(342,92)
\put(30,40){$\po$}\put(40,74){\vector(0,-1){58}}
\put(0,3){$\mathfrak{so}(H,g)\oplus\RR\Id{H}$}
\put(24,80){$\End H$} \put(60,83){\vector(1,0){60}} \put(85,86){$\pi$}
\put(123,80){$\mathfrak{sl}(U)\oplus\RR\Id{U}~\subset~\End U$}
\put(224,75){\vector(-3,-1){174}} \put(130,36){$\imath$}
\put(243,83){\vector(1,0){60}} \put(268,86){$\varkappa$}
\put(306,80){$\End W$}
\end{picture}$$

\subsection{Spinor bundles and connections}
\label{ss:Spinor bundles and connections}

We consider a vector bundle
$S\onto M$ with complex $2$-dimensional  fibers over the real manifold $M$.
For the moment we make no special assumption about $M$, including dimension,
nor we assume any special relation between $S$ and the tangent space $\TO M$:
that relation is mediated by the tetrad, which will be introduced
as a subsequent step in~\sref{ss:Two-spinor tetrad}.

Doing the constructions
of~\sref{ss:Two-spinor algebra and Lorentzian geometry}%
--\ref{ss:Endomorphism decomposition in spinor and Minkowski spaces}
in each fiber we obtain bundles $\LL$, $U$, $H$ and $W$ over $M$,
with smooth natural structures;
a local frame $(\xi_\sA)$ of $S$
yields the associated frames of the other bundles.
Moreover we'll use local coordinates $(\xx^a)$ on $M$.
Note that the fibers of $H$, in particular,
are endowed with a Lorentz metric.

A $\CC$-linear connection $\Cs$ of \hbox{$S\to M$},
called a \emph{$2$-spinor connection},
is expressed by coefficients \hbox{$\Cs\!\iIi{a}{\sA}{\sB}:M\to\CC$}\,.
Their complex conjugates
are the coefficents $\bar\Cs\!\iIi a\cA\cB$ of the induced
\emph{conjugate connection} $\bar\Cs$ of \hbox{$\Sc\to M$},
characterized by the rule \hbox{$\nabla\bar s=\overline{\nabla s}$}.
The components of the induced \emph{dual connection}
$\smash{\ost\Cs}$ of \hbox{$S^*\onto M$}
are \hbox{$\smash{\ost\Cs}\!\iI{a\sB}\sA=-\Cs\!\iIi a\sA\sB$}\,.
A similar relation holds between $\bar\Cs$ and the
\emph{anti-dual connection} $\smash{\ost\Cs}$ of \hbox{$\Sa\onto M$}.
In brief we write \hbox{$\smash{\ost\Cs}{}_{\!a}=-\Cs_{\!a}^*$},
where $\Cs_{\!a}$ is a shorthand for the matrix of the coefficients
and $\Cs_{\!a}^*$ is the transposed matrix; similarly,
we write \hbox{$\smash{\ost{\bar\Cs}}{}_{\!a}=-\bar\Cs{}_{\!a}^*$}\,.

A 2-spinor connection yields linear connections of all bundles
associated with $S$.
If we fix a reference connection $\mathrm{B}$ (a `gauge') then
\hbox{$\Cs\,{-}\,\mathrm{B}$} is a tensor field
valued in \hbox{$\TS M\tn\End S$};
hence, with proper care,
we can describe the relations among the various connections
in terms of bundle endomorphisms using the notions
exposed in~\sref{ss:Endomorphism decomposition in spinor and Minkowski spaces},
with obvious extensions of the needed operations.
In particular:

\myitem~the induced connection of $\weu2S$
is denoted as \hbox{$\hat\Cs\equiv\Tr\Cs$},
with coefficients \hbox{$\hat\Cs\!_a=\Cs\!\iIi a\sA\sA$}\;;

\myitem~the induced connection of $\LL$ has the coefficients
\hbox{$G\!_a\equiv\oh(\Cs\!\iIi a\sA\sA+\bar\Cs\!\iIi a\cA\cA)$},
namely \hbox{$\na_a l=-G\!_a\,l$}\,,
and can be regarded as the ``real part'' of $\hat\Cs$\;;

\myitem~the induced connection of $S\tn\Sc$ is denoted as $\imath\Cs$,
with coefficients
$$(\imath\Cs)\iIi a\AAd\BBd=
\Cs\!\iIi a\sA\sB\,\d\Ii\cA\cB+\d\Ii\sA\sB\,\bar\Cs\!\iIi a\cA\cB~;$$

\myitem~the induced connection of $U$ is denoted as $\tilde\Cs$,
with coefficients
\hbox{$\tilde\Cs\!\iIi a\sA\sB=\Cs\!\iIi a\sA\sB-\oh\,G\!_a\,\d\Ii\sA\sB$}\;;

\myitem~the induced connection of $H$
is denoted as \hbox{$\tilde\G\equiv\imath\tilde\Cs$},
with coefficients
$$\tilde\G\!\iIi a\AAd\BBd=
\tilde\Cs\!\iIi a\sA\sB\,\d\Ii\cA\cB+\d\Ii\sA\sB\,\Bar{\tilde\Cs}\!\iIi a\cA\cB=
(\imath\Cs)\iIi a\AAd\BBd-G\!_a\,\d\Ii\sA\sB\,\d\Ii\cA\cB~.$$
\smallbreak

Above, induced connections are expressed in the frames
induced by $\bigl(\xi_\sA\bigr)$.
Conversely
\begin{align*}
\tilde\Cs\!\iIi a\sA\sB&=\pi(\imath\Cs)\iIi a\sA\sB=
\oh\,(\imath\Cs)\iIi a{\sA\cA}{\sB\cA}
-\oo\,(\imath\Cs)\iIi a\CCd\CCd\,\d\Ii\sA\sB=
\\[6pt]
&=\oh\,\tilde\G\!\iIi a{\sA\cA}{\sB\cA}=\pi(\imath\tilde\Cs)\iIi a\sA\sB~.
\end{align*}
With regard to the latter expression, in particular,
we note that \hbox{$\tilde\G\!\iIi a\AAd\AAd=0$}\,.
Furthermore $\tilde\G$ turns out to be a \emph{metric connection},
preserving the Lorentz fiber structure of $H$;
the notion of torsion, on the other hand,
needs a soldering form, and will be introduced later
(\sref{ss:Two-spinor tetrad}).

It's not difficult to check that similar relations hold
among the curvature tensor $R$ of $\Cs$ and the
curvature tensors of the induced connections. In particular
\begin{align*}
&R\iIi{ab}\AAd\BBd\equiv(\imath R)\iIi{ab}\AAd\BBd=
R\iIi{ab}\sA\sB\,\d\Ii\cA\cB+\d\Ii\sA\sB\,\bar R\iIi{ab}\cA\cB~,
\\[6pt]
&\tilde R\iIi{ab}\sA\sB=R\iIi{ab}\sA\sB+\oh\,\de_{[a}G_{b]}\,\d\Ii\sA\sB=
\\&\phantom{\tilde R\iIi{ab}\sA\sB}
=\oh\,R\iIi{ab}{\sA\cA}{\sB\cA}-\oo\,R\iIi{ab}\CCd\CCd\,\d\Ii\sA\sB=
\oh\,\tilde R\iIi{ab}{\sA\cA}{\sB\cA}~.
\end{align*}

We also consider the induced connection $\varkappa(\tilde\Cs)$
on the 4-spinor bundle \hbox{$W\equiv U\oplus\Ua$}.
This can be then expressed in terms of the components of the Dirac map as
$$\varkappa(\tilde\Cs)=
\oo\tilde\G\!\iIi a\lambda\mu\,
(\gamma_\lambda\gamma^\mu-\gamma^\mu\gamma_\lambda)~,$$
since \hbox{$\tilde\G\!\iIi a\lambda\lambda=0$}\,,
where the components of $\tilde\G$ are now expressed
in the Pauli frame $\bigl(\tau_\lambda\bigr)$.

\bigbreak
Last but not least we consider the induced connection $Y$ of $\weu2U$,
whose fibers (\sref{ss:Two-spinor algebra and Lorentzian geometry})
have a natural Hermitian structure.
Indeed $Y$ preserves that structure,
and its coefficients can be written as $\iO Y\!_a$ where
\hbox{$Y\!_a=\tfrac1{2\iO}(\Cs\!\iIi a\sA\sA-\bar\Cs\!\iIi a\cA\cA)$}
is the imaginary part of $\hat\Cs\!_a$\,;
namely \hbox{$\hat\Cs\!_a=G\!_a+\iO Y\!_a$}\,.
In particular, we get \hbox{$\na_a\e=\iO Y\!_a\,\e$}\,.

\subsection{Breaking of dilaton symmetry}
\label{ss:Breaking of dilaton symmetry}

In a general theory of fields that are sections of the various bundles
derived from $S$ one has to allow a ``dilaton'' field \hbox{$M\to\LL$}\,.
In the literature this issue has been considered
under various angles~\cite{AMS14,CorianoDRQS13,Fa09,FKD10,HCMN95,
LavelleMcMullan95,NoBi12,ILM10,Ohanian16,Pervushin_etal12,Pons09,RySh10},
though a conclusive approach seems to be still lacking.
One intriguing possibility is that the dilaton be closely related
to the Higgs field.
My own ideas about such speculations
have been expressed in two papers~\cite{C10a,C15b}.

In the sequel, however, we'll work for simplicity in a ``conservative'' setup
in which the dilaton symmetry is broken by some mechanism
we do not worry about here.
This enables a formulation,
sketched in~\sref{ss:Einstein-Cartan-Maxwell-Dirac fields},
of a theory of Einstein-Cartan-Maxwell-Dirac fields
which is based on geometric construction using $S$ with the only added
assumption of such symmetry breaking.

A weak form of this assumption, sufficient for our present purposes,
can be expressed as the requirement that the connection $G$ induced on $\LL$
has vanishing curvature, that is \hbox{$\dO G=0$}\,.
Hence one can always find local charts such that $G_a=0$\,,
and this amounts to gauging away the conformal symmetry.

In practice we may wish to simplify certain arguments
by making the stronger assumption that the bundle \hbox{$\LL\onto M$}
be trivial, that is a global product.
This means that we actually regard $\LL$ just a semi-vector space,
the space of length units.
In a ``natural unit setting'',
coupling constants now arise as elements in $\LL$
(see also footnote~\ref{footnote:unitspaces}
on page~\pageref{footnote:unitspaces}).

\subsection{Two-spinor tetrad}
\label{ss:Two-spinor tetrad}

Henceforth we assume that $M$ is a real $4$-dimensional manifold,
and consider a section \hbox{$\Th:M\to\LL\tn H\tn\TS M$}.
Note that $\Th$ can be seen as a linear morphism \hbox{$\TO M\to\LL\tn H$},
and, if it is non-degenerate, as a `scaled' \emph{tetrad}.
We write its coordinate expression as
$$ \Th=\Th_a^\lambda\,\tau_\lambda\tn\dx^a=
\Th_a^\AAd\,\zeta_\sA\tn\bar\zeta_\cA\tn\dx^a~,$$
where the coefficients $\Th_a^\lambda$ and $\Th_a^\AAd$ are $\LL$-valued
(namely have the physical dimensions of a \emph{length}).

Given a tetrad, the geometric structure of the  fibers of $H$
yields a similar, \emph{scaled} structure on the  fibers of $\TO M$.
Namely if we now denote by $\tilde g$\,, $\tilde\eta$ and $\tilde\gamma$
the Lorentz metric,
the $\tilde g$-normalized volume form and the Dirac map of $H$,
we get similar spacetime objects
\begin{align*}
g&\equiv\Th^*\tilde g=
\tilde g_{\lambda\mu}\,\Th_a^\lambda\,\Th_b^\mu\,\dx^a\tn\dx^b
=\e_{\sA\sB}\e_{\cA\cB}\,\Th_a^\AAd\,\Th_b^\BBd\,\dx^a\tn\dx^b~,
\\[6pt]
\eta&\equiv\Th^*\tilde\eta=\det\Th\,\dO^4\xx\equiv
\det\Th\,\dx^0\we\dx^1\we\dx^2\we\dx^3~,
\\[6pt]
\gamma&\equiv\tilde\gamma\comp\Th=
\Th_a^\lambda\,\gamma_\lambda\tn\dx^a\equiv\gamma_a\tn\dx^a~.
\end{align*}
These relations show that we can regard $\Th$ as a sort of
``square root of the metric''.

The inverse morphism can be obtained by moving indices via the metric, namely
$$\cev\Th{}_\mu^b=\Th_\mu^b\equiv\Th_a^\lambda\,\tilde g_{\lambda\mu}\,g^{ab}~.$$
Moreover a Pauli frame yields the orthonormal scaled spacetime frame and dual frame
$$\bigl(\Th_\lambda\bigr)\equiv\cev\Th(\tau_\lambda)=
\bigl(\Th^a_\lambda\,\de\xx_a\bigr)~,\qquad
\bigl(\Th^\lambda\bigr)\equiv\Th^*(\tau^\lambda)=
\bigl(\Th_a^\lambda\,\dx^a\bigr)~.$$

If $\Cs$ is a 2-spinor connection
then a non-degenerate tetrad $\Th:\TO M\to\LL\tn H$ yields a unique
connection $\G$ on $\TO M$, characterized by the condition that $\Th$
be covariantly constant with respect to the couple $(\G,\tilde\G)$.
Indeed the condition \hbox{$\nabla\Theta=0$} reads
$$\de_b\Th_a^\lambda+\G\!\iIi bca\,\Th_c^\lambda
-\tilde\G\!\iIi b\lambda\mu\,\Th_a^\mu=0~,$$
while the components of $\G$ in the orthonormal frame $\bigl(\Th_\lambda\bigr)$
coincide with the components of $\tilde\G$ in the associated Pauli frame:
\hbox{$\G\!\iIi a\lambda\mu=\tilde\G\!\iIi a\lambda\mu$}.
Moreover $\G$ is metric, \hbox{$\nabla[\G]g=0$}.
The curvature tensors of $\G$ and $\tilde\G$ are similarly related by
$R\iIi{ab}{\lambda}{\mu}=\tilde R\iIi{ab}{\lambda}{\mu}$\,, that is
$$R\iIi{ab}{c}{d}=\tilde R\iIi{ab}{\lambda}{\mu}\,\Th_\lambda^c\,\Th_d^\mu~.$$
The Ricci tensor and the scalar curvature can be expressed as
$$R_{ad}=R\iIi{ab}{b}{d}=
\tilde R\iIi{ab}{\lambda}{\mu}\,\Th_\lambda^b\,\Th_d^\mu~,\qquad
R\iI{a}{a}=\tilde R\iI{ab}{\lambda\mu}\,\Th_\lambda^b\,\Th_\mu^a~.$$

In general $\G$ will have non-vanishing torsion,
which can be expressed as
$$\Th_c^\lambda\,T\Ii{c}{ab}=\de_{[a}^{\phantom{a}}\Th_{b]}^\lambda
+\Th_{[a}^\mu\,\tilde\G\!\iIi{b]}{\lambda}{\mu}~.$$

\subsection{Einstein-Cartan-Maxwell-Dirac fields}
\label{ss:Einstein-Cartan-Maxwell-Dirac fields}

The field theory we are going to sketch,
as presented in previous papers~\cite{C98, C00b, C07},
is based on ``minimal geometric data''
in the sense that the unique such datum is a vector bundle \hbox{$S\onto M$},
with complex 2-dimensional fibers and real 4-dimensional base manifold.
The basic idea is to assume no further background structure:
all other bundles and fixed geometric objects
are derived from $S$ using only geometrical constructions.
Any needed bundle section which is not determined by $S$
is assumed to be a field.
A natural Lagrangian can then be written, yielding a field theory
which turns out to be essentially equivalent to a classical theory
of Einstein-Cartan-Maxwell-Dirac fields.
Admittedly, however, there is \emph{one} further assumption that must be made
at this level,
namely the breaking of dilatonic symmetry
as described in~\sref{ss:Breaking of dilaton symmetry};
without that we deal with a more general theory.
Accordingly we regard $\LL$ as a fixed semi-vector space,
whose unique role consists of taking care of physical dimensions
in a ``natural'' unit setting (\hbox{$\hbar=c=1$}).

The fields are taken to be the tetrad $\Th$\,,
the $2$-spinor connection $\Cs$,
the electromagnetic field $F$ and the electron field $\psi$\,.
The gravitational field is represented both by $\Th$
and $\tilde\Cs$, the latter being regarded as the gravitational part of $\Cs$.
If $\Th$ is non-degenerate then one obtains,
as in the standard metric-affine approach~\cite{HCMN95,FK82},
essentially the Einstein equation and the equation for torsion;
the metricity of the spacetime connection is a further consequence.
The theory, however, is non-singular also if $\Th$ is degenerate.
Also note that in this approach
the spacetime metric $g$ and the spacetime connection $\G$
are not independent fields, but rather byproducts of the formalism.
A particular consequence of this fact is that we cannot just require
the torsion to vanish.

The Dirac field \hbox{$\psi\equiv(u,\chi)$}
is a section of the Dirac bundle $W$ with physical dimensions $\LL^{-3/2}$,
and is assumed to represent a particle with one-half spin,
mass \hbox{$m\in\LL^{-1}$} and charge \hbox{$q\in\RR$}\,.

We assign the role of the electromagnetic potential to another sector of $\Cs$,
namely the induced Hermitian connection $Y$ of $\weu2U$
whose coefficients we denote as $\iO Y\!_a$
(locally one also writes \hbox{$Y\!_a\equiv qA_a$}\,,
where $A$ is a local 1-form).

The electromagnetic field is represented by a spacetime 2-form $F$ or, equivalently,
by a section \hbox{$\tilde F:M\to\LL^{-2}\tn\weu2H^*$}
related to it by \hbox{$F\equiv\Th^*\tilde F$}.
The relation between $Y$ and $F$ follows as one of the field equations.

The total Lagrangian density
\hbox{$\mathcal{L}=(\ell\spec{grav}+ \ell\spec{em}+ \ell\spec{Dir})\,\dO^4\xx$}
is the sum of gravitational, electromagnetic and Dirac terms.
These can be written in coordinate-free form,
but the coordinate expressions are perhaps more readable
without special explanations. We have
\begin{align*}
\ell\spec{grav}&=\tfrac{1}{8\Bbbk}\,\e_{\lambda\mu\nu\rho}\,\e^{abcd}\,
\tilde R\iI{ab}{\lambda\mu}\,\Th_c^\nu\,\Th_d^\rho~,
\\[6pt]
\ell\spec{em}&=
-\oq\,\e^{abcd}\,\e_{\lambda\mu\nu\rho}\,
\de_aY\!_b\,\tilde F^{\lambda\mu}\,\Th_c^\nu\Th_d^\rho
+\oq\,\tilde F^{\lambda\mu}\tilde F_{\lambda\mu}\,\det\Th~,
\\[6pt]
\ell\spec{Dir}&=
\isq\,\breve\Th^a_{\AAd}\,
\Bigl(\na_au^\sA\,\bar u^\cA-u^\sA\,\na_a\bar u^\cA
+\e^{\sA\sB}\bar\e^{\cA\cB}(\bar\chi_\sB\,\na_a\chi_\cB
-\na_a\bar\chi_\sB\,\chi_\cB\,)\Bigr)
\\
&\hspace{7cm}-m\,(\bar\chi_\sA u^\sA+\chi_\cA\,\bar u^\cA\,)\,\det\Th~,
\end{align*}
where
$$\breve\Th^a_{\AAd}\equiv\osq\,\sigma\Ii\lambda{\AAd}\,\breve\Th^a_\lambda
\equiv
\osq\,\sigma\Ii\lambda{\AAd}\,\bigl(
\tfrac{1}{3!}\,\e^{abcd}\,\e_{\lambda\mu\nu\rho}\,
\Th_b^\mu\Th_c^\nu\Th_d^\rho\bigr)$$
and $\Bbbk$ stands for Newton's gravitational constant.

The main results obtained by
writing down the Euler-Lagrange equations deriving from $\mathcal{L}$
can be summarized as follows.

\myitem~The $\Th$-component corresponds
(in the non-degenerate case) to the Einstein equations.

\myitem~The $\tilde\G$-component gives the equation for torsion.
Hence one sees that the spinor field is a source for torsion,
and that in this context a possible torsion-free theory is not natural.

\myitem~The $F$-component reads \hbox{$F=2\,\dO Y$} in the non-degenerate case,
and of course this yields the first Maxwell equation \hbox{$\dO F=0$}\,.

\myitem~The $Y$-component reduces, in the non-degenerate case,
to the second Maxwell equation \hbox{$\oh\,{*}\dO{*}F=j$}\,,
where the 1-form $j$ is the \emph{Dirac current}.

\myitem~The $\bar u$- and $\bar\chi$-components yield the \emph{Dirac equation}
\hbox{$(\iO\,\nasl-m+\tfrac{\iO}{2}\,\gamma^aT\Ii{b}{ab})\psi=0$}
for \hbox{$\psi\equiv(u,\chi)$}\,.

\myitem~The $u$- and $\chi$-components yield the Dirac equation
for the Dirac adjoint \hbox{$\bar\psi=(\bar\chi,\bar u)$}\,.

\smallbreak

\section{Lie derivatives in spinor geometry and tetrad gravity}
\label{s:Lie derivatives in spinor geometry and tetrad gravity}

\subsection{Lie derivative of spinors}
\label{ss:Lie derivative of spinors}

We start by looking for a natural definition
of Lie derivative of sections \hbox{$w:M\to H$}
with respect to a vector field \hbox{$\sX:M\to\TO M$}.
We observe that $\cev\Th w$ is a vector field on $M$,
and so is the Lie bracket \hbox{$[\sX,\cev\Th w]\equiv\LO_\ssX(\cev\Th w)$}\,.
Then we obtain the section \hbox{$\Th[\sX,\cev\Th w]:M\to H$}.
By a straightforward calculation
this is found to have the coordinate expression
\begin{align*}
&(\Th[\sX,\cev\Th w])^\lambda=\na_\ssX w^\lambda-\Xi\Ii\lambda\mu\,w^\mu~,
\\[6pt]
&\Xi\Ii\lambda\mu\equiv
(\na_a\sX^b+T\Ii b{ca}\,\sX^c)\,\cev\Th{}_\mu^a\,\Th_b^\lambda~,
\end{align*}
where $\na_\ssX w$ is the covariant derivative
with respect to the connection $\tilde\G$
(\sref{ss:Spinor bundles and connections}).

We now face the following issue:
in order to recover $\Th[\sX,\cev\Th w]$ from an operation
performed on 2-spinors by means of the Leibnitz rule,
\hbox{$\Xi:M\to\End H$} must be valued
in \hbox{$\mathfrak{so}(H,g)\oplus\RR\Id{H}$}
(\sref{ss:Endomorphism decomposition in spinor and Minkowski spaces}).
Since this is not true in general, we consider a modified operation $\LO_\ssX$
acting on sections \hbox{$w:M\to H$}
$$\LO_\ssX w:=\na_\ssX w-(\po\Xi)w$$
in which $\Xi$ is replaced by its projection onto
\hbox{$\mathfrak{so}(H,g)\oplus\RR\Id{H}$}\,;
in term of components, $\Xi\Ii\lambda\mu$ is replaced by
$$\hat\Xi\Ii\lambda\mu\equiv(\po\Xi)\Ii\lambda\mu=
\oh\,\bigl(\Xi\Ii\lambda\mu\,{-}\,(\Xi^\dagger)\Ii\lambda\mu\bigr)
+\oq\,\Xi\Ii\nu\nu\,\d\Ii\lambda\mu~.$$

Then indeed by setting
\begin{align*}
&\xi\equiv\pi(\Xi)=\pi(\hat\Xi):M\to\End U
\\[6pt]
&\LO_\ssX u=\na_\ssX u-\xi(u)~,&&
\LO_\ssX\bar u\equiv\overline{\LO_\ssX u}=\na_\ssX\bar u-\bar\xi(\bar u)~,
\\[6pt]
&\LO_\ssX\chi=\na_\ssX\chi+\bar\xi^*\chi~,&&
\LO_\ssX\bar\chi\equiv\overline{\LO_\ssX\chi}=\na_\ssX\bar\chi+\xi^*\bar\chi~,
\end{align*}
where \hbox{$u:M\to U$}, \hbox{$\chi:M\to\Ua$},
recalling the definitions and results
in~\sref{ss:Endomorphism decomposition in spinor and Minkowski spaces}
we easily check that all the natural Leibnitz rules are fulfilled.
In particular we find
$$\LO_\ssX(u\tn\bar u)=(\LO_\ssX u)\tn\bar u+u\tn\LO_\ssX\bar u~,\qquad
\sX.\bang{\bar\chi,u}=\bang{\LO_\ssX\bar\chi,u}+\bang{\bar\chi,\LO_\ssX u}~.$$
We have the coordinate expression
$$\LO_\ssX u^\sA=
\sX^a\,(\de_au^\sA-\Cs\iIi a\sA\sB\,u^\sB)-\xi\Ii\sA\sB\,u^\sB~,
\qquad
\xi\Ii\sA\sB=\oh\,\Xi\Ii{\sA\cA}{\sB\cA}-\oo\,\Xi\Ii\CCd\CCd\,\d\Ii\sA\sB~.$$

The Lie derivative of the 4-spinor
\hbox{$\psi\equiv(u,\chi):M\to W\equiv U\oplus\Ua$} is then
\begin{align*}
\LO_\ssX\psi&=\na_\ssX\psi-\bigl(\xi\,,-\bar\xi{}^*\bigr)\psi=
\na_\ssX\psi-\varkappa(\pi\Xi)=
\\[6pt]
&=\na_\ssX\psi
-\oo\,\Xi_{\lambda\mu}\,(\gamma^\lambda\gamma^\mu-\gamma^\mu\gamma^\lambda)\psi
+\oo\,\Xi\Ii\nu\nu\,\gamma_5\psi~.
\end{align*}

When the torsion vanishes and $\sX$ is a Killing vector field
(then \hbox{$\Xi\Ii\nu\nu=0$})
one essentially gets the usual Lie derivative of Dirac spinors~\cite{Kosmann71},
though a careful reader may notice an opposite sign in the second term.
The standard expression can be recovered
by exchanging the roles of the bundles $U$ and $\Ua$,
so that the difference can be eventually ascribed to conventions
affecting representations of the involved Lie algebras.
Similarly one sees that our expression for $\LO_\ssX u^\sA$
is the same as that in Penrose-Rindler~\cite{PR88},~\S6.6,
when the torsion vanishes and $\sX$ is a conformal Killing vector field.

\remark~The spacetime connection being a necessarily ingredient
in the Lie derivative of spinors is a consequence of \hbox{$\nabla\Th=0$}\,.
Also note that the possible non-vanishing of the torsion
implies that the condition \hbox{$\Xi=\po\Xi$} is not equivalent
to the requirement that $\sX$ be a conformal Killing vector field.
\smallbreak

\remark~The Fermi transport of spinors
can be introduced by an analogous construction~\cite{C09a}
starting from the Fermi transport of world-vectors.
\smallbreak

\subsection{Lie derivative of the tetrad}
\label{ss:Lie derivative of the tetrad}

Though the notion of Lie derivative of spinors proposed
in~\sref{ss:Lie derivative of spinors}
is well-defined for any vector field $\sX$,
it is actually independent of the symmetric trace-free part of
\hbox{$\Xi\equiv\nabla\sX+\sX\pint T$}.
However that part has not merely disappeared from view,
but is related to a natural definition of Lie derivative of the tetrad.
This follows from imposing the Leibnitz rule
$$\LO_\ssX(\cev\Th w)=(\LO_\ssX\cev\Th)w+\cev\Th\LO_\ssX w~,\qquad
w:M\to H~,$$
whence
\begin{align*}
\Th\bigl((\LO_\ssX\cev\Th)w\bigr)&=
\Th\bigl(\LO_\ssX(\cev\Th w)\bigr)-\LO_\ssX w=
\na_\ssX w-\Xi(w)-\bigl(\na_\ssX w-\po\Xi(w)\bigr)=
\\[6pt]
&=\bigl(\po\Xi-\Xi\bigr)(w)~.
\end{align*}

Requiring now \hbox{$\LO_\ssX(\Th\comp\cev\Th)=0$} we eventually get
\begin{align*}
&\Th\comp\LO_\ssX\cev\Th=\po\Xi-\Xi~,\qquad
\LO_\ssX\Th\comp\cev\Th=\Xi-\po\Xi~,
\\[6pt]\Rightarrow\quad
&\LO_\ssX\Th=(\Xi-\po\Xi)\comp\Th~,
\end{align*}
that is
$$\LO_\ssX\Th_a^\lambda=(\Xi\Ii\lambda\mu-\hat\Xi\Ii\lambda\mu)\,\Th^\mu_a=
\oh\,(\Xi\Ii\lambda\mu+\Xi\iI\mu\lambda)\,\Th^\mu_a-\oq\,\Xi\Ii\nu\nu\,\Th^\lambda_a~.$$

One may wonder why the above expression does not contain
the derivatives of the components of $\Th$\,.
The answer is that they are actually contained in the torsion,
which is contained in $\Xi$\,.
In fact we can recover our result by a straightforward
coordinate calculation from
$$\LO_\ssX\Th_a^\lambda=\sX^b\,\de_b\Th_a^\lambda+\Th_b^\lambda\,\de_a\sX^b
-\sX^b\,\tilde\G\!\iIi b\lambda\mu\,\Th_a^\mu-\hat\Xi\Ii\lambda\mu\,\Th_a^\mu$$
and then using
\hbox{$\de_b\Th_a^\lambda=
\tilde\G\!\iIi b\lambda\mu\,\Th_a^\mu-\G\!\iIi bca\,\Th_c^\lambda$}
which is the coordinate expression of \hbox{$\nabla\Th=0$}\,.

\subsection{Lie derivative of the spinor connection}
\label{ss:Lie derivative of the spinor connection}

If $\G$ is an arbitrary linear connection of the tangent bundle of $M$,
then its Lie derivative along a vector field $\sX$ is the tensor field
\hbox{$\LO_\ssX\G:M\to\TS M\tn\TO M\tn\TS M$}
characterized by the identity~\cite{Yano55}
$$\LO_\ssX\G\pint\sZ=\nabla\LO_\ssX\sZ-\LO_\ssX\nabla\sZ$$
holding for any vector field $\sZ$\,.
Its coordinate expression turns out to be
\begin{align*}
\LO_\ssX\G\!\iIi abc&=-\na_a\na_c\ssX^b-\na_a(\ssX^d\,T\Ii b{dc})
+\ssX^d R\iIi{ad}bc=
\\[6pt]
&=-\na_a\Xi\Ii bc+\ssX^d R\iIi{ad}bc~.
\end{align*}
This notion can be applied in particular to the Riemannian spacetime connection,
and as such it appears in the literature mainly in considerations
related to energy tensors~\cite{KijowskiGRG97,Padamnabhan1312.3253},
possibly in a somewhat disguised form~\cite{LandauLifchitz68,HE73}.

The Lie derivative of the linear connection $\tilde\G$ of \hbox{$H\onto M$}
can be obtained by extending that construction.
We set
$$\LO_\ssX\tilde\G:M\to\TS M\tn\End H:
w\mapsto\nabla\LO_\ssX w-\LO_\ssX\nabla w~,$$
where \hbox{$w:M\to H$} and $\LO_\ssX w$ is the operation
introduced in~\sref{ss:Lie derivative of spinors}.
A calculation then yields
$$\LO_\ssX\tilde\G\!\iIi a\lambda\mu=
-\na_a\hat\Xi\Ii\lambda\mu+\sX^d\,\tilde R\iIi{ad}\lambda\mu~.$$

\remark~For an arbitrary vector field $\sX$ we have
\hbox{$\LO_\ssX\tilde\G\!\iIi a\lambda\lambda=
-\na_a\hat\Xi\Ii\lambda\lambda\neq0$}\,,
so that the ``deformed connection'' \hbox{$\tilde\G+\LO_\ssX\tilde\G$}
needs not be metric.\smallbreak

Similarly, the Lie derivative
\hbox{$\LO_\ssX u^\sA=\na_\ssX u^\sA-\xi\Ii\sA\sB\,u^\sB$}
of spinors \hbox{$u:M\to U$} yields the Lie derivative
of the linear connection $\Cs$ of \hbox{$U\onto M$} as
$$\LO_\ssX\Cs:M\to\TS M\tn\End U~,\qquad
\LO_\ssX\Cs\pint u=\nabla\LO_\ssX u-\LO_\ssX\nabla u~.$$
Again, a straightforward calculation yields
$$\LO_\ssX\Cs\iIi a\sA\sB=-\na_a\xi\Ii\sA\sB+\sX^d\,R\iIi{ad}\sA\sB~.$$

Now, using the relation between $\Xi$ and $\xi$
and the analogous relation (\sref{ss:Spinor bundles and connections})
between $R[\Cs]$ and \hbox{$\tilde R\equiv R[\tilde\G]$},
it's not difficult to check that also
$$\LO_\ssX\Cs=\pi(\LO_\ssX\tilde\G)~,\qquad
\LO_\ssX\tilde\G=\imath(\LO_\ssX\Cs)~.$$
In coordinates, these read
\begin{align*}
&\LO_\ssX\Cs\iIi a\sA\sB=
\oh\,\LO_\ssX\tilde\G\!\iIi a{\sA\cA}{\sB\cA}
-\oo\,\LO_\ssX\tilde\G\!\iIi a\CCd\CCd\,\d\Ii\sA\sB~,
\\[6pt]
&\LO_\ssX\tilde\G\!\iIi a\AAd\BBd=
\LO_\ssX\Cs\iIi a\sA\sB\,\d\Ii\cA\cB+\d\Ii\sA\sB\,\LO_\ssX\bar\Cs\iIi a\cA\cB~.
\end{align*}

We recall (\sref{ss:Spinor bundles and connections}) that $\Cs$ yields
connections of $\Uc$, $U^*$ and $\Ua$.
The Lie derivatives of all these are naturally defined
by straightforward extensions of the above procedure,
and their coordinate expressions are easily checked to be
in the same mutual relations.
Moreover we get the Lie derivative of the 4-spinor connection,
which can be expressed as
$$\LO_\ssX(\Cs,\ost{\bar\Cs})_a=
(\LO_\ssX\Cs_a,\LO_\ssX\ost{\bar\Cs}_a)=
\oo\,\LO_\ssX\tilde\G\!\iIi a\lambda\mu\,
(\gamma_\lambda\gamma^\mu-\gamma^\mu\gamma_\lambda)
+\oo\,\LO_\ssX\tilde\G\!\iIi a\lambda\lambda\,\gamma_5~,$$
where \hbox{$\LO_\ssX\tilde\G\!\iIi a\lambda\lambda=
-\na_a\hat\Xi\Ii\lambda\lambda$}\,.

Our notion of Lie derivative of 2-spinors naturally yields
the Lie derivatives of the curvature tensors of $\Cs$ and $\tilde\G$.
We obtain the coordinate expressions
\begin{align*}
\LO_\ssX R\iIi{ab}\sA\sB&=\sX^c\,\de_cR\iIi{ab}\sA\sB
+\de_a\sX^c\,R\iIi{cb}\sA\sB+\de_b\sX^c\,R\iIi{ac}\sA\sB
+[R_{ab}\,,\sX^a\Cs\!_a\,{+}\,\xi\,]\Ii\sA\sB~,
\\[6pt]
\LO_\ssX\tilde R\iIi{ab}\lambda\mu&=\sX^c\,\de_c\tilde R\iIi{ab}\lambda\mu
+\de_a\sX^c\,\tilde R\iIi{cb}\lambda\mu+\de_b\sX^c\,\tilde R\iIi{ac}\lambda\mu
+[\tilde R_{ab}\,,\sX^a\tilde\G\!_a\,{+}\,\hat\Xi]\Ii\lambda\mu~,
\end{align*}
where
\hbox{$[R_{ab}\,,\xi\,]\Ii\sA\sB\equiv
R\iIi{ab}\sA\sC\,\xi\Ii\sC\sB-\xi\Ii\sA\sC\,R\iIi{ab}\sC\sB$}
and the like
(brackets denote commutators of fiber endomorphisms).

Then it is not difficult to check that the algebraic relation
between these two objects is essentially the same
as the relation between $\Cs$ and $\tilde\G$.
Moreover let us regard \hbox{$\Cs'\equiv\Cs+\LO_\ssX\Cs$} as a
``deformed'' spinor connection;
then its curvature tensor turns out to be \hbox{$R'=R+\LO_\ssX R$}
up to terms which are of second order in the Lie derivatives.
A similar statement holds true for the curvature of the
deformed connection \hbox{$\tilde\G'\equiv\tilde\G+\LO_\ssX\tilde\G$}.

\remark~For the reader who is familiar
with the Fr\"olicher-Nijenhuis bracket
of tangent-valued forms~\cite{FN56,MaMo83a,Mi01},
we can recast above results in a convenient way.
We first observe that if \hbox{$E\onto M$} is any vector bundle
then an $\End\!E$-valued $r$-form \hbox{$M\to\weu{r}\TS M\tn\End\!E$}
can be regarded as a vertical-valued form on $E$.
A linear connection can also be regarded as a tangent-valued 1-form,
and its curvature tensor as a vertical-valued 2-form.
Moreover a vector field on $E$ is a tangent-valued 0-form.
In particular, both $\xi$ and
$$\sX\pint\Cs=\sX^a\,\de\xx_a+\sX^a\,\Cs\iIi a\sA\sB\,\zeta^\sB\,\zeta_\sA$$
are vector fields on $U$.
Indeed, the latter is the \emph{horizontal prolongation} of $\sX$
through the connection $\Cs$.
Similarly, \hbox{$\hat\Xi\equiv\po\Xi$} and \hbox{$\sX\pint\tilde\G$}
are vector fields on $H$.
A computation then yields
\begin{align*}
&\LO_\ssX\Cs=\fnb{\,\sX\pint\Cs+\xi\,,\,\Cs\,}~,
&& \LO_\ssX R=\fnb{\,\sX\pint\Cs+\xi\,,\,R\,}~,
\\[6pt]
&\LO_\ssX\tilde\G=\fnb{\,\sX\pint\tilde\G+\hat\Xi\,,\,\tilde\G\,}~,
&&\LO_\ssX\tilde R=\fnb{\,\sX\pint\tilde\G+\hat\Xi\,,\,\tilde R\,}~.
\end{align*}
Furthermore $\Th$ can be regarded as a vertical valued 1-form on $H$,
while a 2-spinor \hbox{$u:M\to U$}
can be regarded as a section \hbox{$U\to\VO U$}.
Then we also find
$$\LO_\ssX u=\fnb{\,\sX\pint\Cs+\xi\,,\,u\,}~,\qquad
\LO_\ssX\Th=\fnb{\,\sX\pint\tilde\G+\hat\Xi\,,\,\Th\,}~.$$
\smallbreak

\subsection{Deformed tetrad gravity}
\label{ss:Deformed tetrad gravity}

Consider arbitrarily deformed objects
\hbox{$\G'\equiv\G+\sDe\G$}, \hbox{$\tilde\G'\equiv\tilde\G+\sDe\tilde\G$},
\hbox{$\Th'\equiv\Th+\sDe\Th$}\,.
Then up to second-order terms in the deformations we get
\begin{align*}
\nabla'_{\!c}\Th'{}_a^\lambda&=
\de_c\Th'{}_a^\lambda+\G'\iIi cba\,\Th'{}_b^\lambda
-\tilde\G'\iIi c\lambda\mu\,\Th'{}_a^\mu=
\\[6pt]
&=\de_c(\Th_a^\lambda+\sDe\Th_a^\lambda)
+(\G\!\iIi cba+\sDe\G\!\iIi cba)\,(\Th_b^\lambda+\sDe\Th_b^\lambda)
-(\tilde\G\!\iIi c\lambda\mu+\sDe\tilde\G\!\iIi c\lambda\mu)\,
(\Th_a^\mu+\sDe\Th_a^\mu)=
\\[6pt]
&=\na_c\Th_a^\lambda+\na_c(\sDe\Th)_a^\lambda
+\sDe\G\!\iIi cba\,\Th_b^\lambda
-\sDe\tilde\G\!\iIi c\lambda\mu\,\Th_a^\mu+o(\sDe)~.
\end{align*}
Since \hbox{$\nabla\Th=0$}\,, the above relation can be written,
dropping second-order terms in the deformations, as
$$\nabla'_{\!c}\Th'{}_a^\lambda=
\na_c(\sDe\Th)_a^\lambda+\sDe\G\!\iIi cba\,\Th_b^\lambda
-\sDe\tilde\G\!\iIi c\lambda\mu\,\Th_a^\mu~.$$

Now we consider the special case when the deformations are
Lie derivatives along a vector field \hbox{$\sX:M\to\TO M$}, namely
\begin{align*}
\sDe\G\!\iIi cab&\equiv\LO_\ssX\G\!\iIi cab=
-\na_c\Xi\Ii ba+\sX^d\,R\iIi{cd}ba~,
\\[6pt]
\sDe\tilde\G\!\iIi c\lambda\mu&\equiv\LO_\ssX\tilde\G\!\iIi c\lambda\mu=
-\na_c\hat\Xi\Ii\lambda\mu+\sX^d\,R\iIi{cd}\lambda\mu~,
\\[6pt]
\sDe\Th_a^\lambda&\equiv\LO_\ssX\Th_a^\lambda=
\Th_a^\mu\,(\Xi-\hat\Xi)\Ii\lambda\mu~.
\end{align*}
Then we obtain
\begin{align*}
\nabla'_{\!c}\Th'{}_a^\lambda&=\na_c(\Xi-\hat\Xi)\Ii\lambda\mu\,\Th_a^\mu
+(\sX^d\,R\iIi{cd}ba-\na_c\Xi\Ii ba)\,\Th_b^\lambda
+(\na_c\hat\Xi\Ii\lambda\mu-\sX^d\,R\iIi{cd}\lambda\mu)\,\Th_a^\mu=
\\[6pt]
&=\sX^d\,(R\iIi{cd}ba\,\Th_b^\lambda-\sX^d\,R\iIi{cd}\lambda\mu\,\Th_a^\mu)=0~,
\end{align*}
so that the deformed tetrad $\Th'$ is covariantly constant
with respect to the deformed connections $\G'$ and $\tilde\G'$.

In the gravitational field theory formulation
sketched in~\sref{ss:Einstein-Cartan-Maxwell-Dirac fields}
the gravitational field is represented by the couple $(\Th,\Cs)$
while the spacetime connection $\G$ is a byproduct,
characterized by the condition \hbox{$\nabla\Th=0$}\,.
Hence the above result can be interpreted as saying that a deformed couple
$(\Th',\Cs')$ yields the deformed spacetime connection
\hbox{$\G'\equiv\G+\LO_\ssX\G$},
where the deformation is the Lie derivative of $\G$ in the usual sense.

\subsection{A remark on possible extensions}
\label{ss:A remark on possible extensions}

The various connections induced by a 2-spinor connection $\Cs$
on the bundles constructed from $S$ can be regarded as ``pieces''
into which $\Cs$ can be naturally decomposed.
In particular, the imaginary part $\iO Y$ of $\hat\Cs$
is the induced \emph{Hermitian connection} of $\weu2U$.

We note that $Y\!_a$ does not enter
$\hat\Xi\Ii\lambda\mu$ nor $\xi\Ii\sA\sB$\,,
hence its contribution to the Lie derivatives of spinors,
and the other related Lie derivatives,
is limited to the covariant derivative $\na_\ssX$\,.
We may say that the internal geometry of $\weu2U$
is not soldered to spacetime geometry.
An analogous result was found~\cite{C09a} in the construction
of the Fermi transport of spinors along a timelike line.
Actually $Y$ is related to the electromagnetic potential and,
in pure electrodynamics, can be just interpreted as such.

Adding further internal degrees of freedom
means considering new vector bundles, say \hbox{$F\onto M$},
whose fibers are not soldered to spacetime geometry,
and taking tensor products such as $U\tn F$.
In general, in such enlarged setting,
one has no well-defined notion of Lie derivatives
of matter fields and gauge fields
with respect to vector fields on the base manifold $M$.
On the other hand, the notion of Lie derivative with respect to a vector field
on the total manifold is well-defined,
and an important tool in Lagrangian field theory\,---\,%
with particular regard to symmetries.


%
\end{document}